\documentstyle[bezier,epsfig,12pt,preprint,tighten,aps]{revtex}
\begin{document}

\draft

\title{\rightline{{\tt (August 1998)}}
\rightline{{\tt UM-P-98/41}}
\rightline{{\tt RCHEP-98/08}}
\ \\
Do the SuperKamiokande atmospheric neutrino results explain\\
electric charge quantisation?}
\author{R. Foot and R. R. Volkas}
\address{School of Physics\\
Research Centre for High Energy Physics\\
The University of Melbourne\\
Parkville 3052 Australia\\
(foot@physics.unimelb.edu.au, r.volkas@physics.unimelb.edu.au)}

\maketitle

\begin{abstract} 

We show that the SuperKamiokande atmospheric neutrino results 
explain electric charge
quantisation, provided that the oscillation mode 
is $\nu_{\mu} \to \nu_{\tau}$
and that the neutrino mass is of the Majorana type.

\end{abstract}

\newpage

The up-down asymmetry observed by SuperKamiokande\cite{sk} for the 
atmospheric muon-neutrino flux
provides extremely clear evidence for $\nu_{\mu}$ 
disappearence, and hence for the
violation of muon-type lepton number $L_{\mu}$. 
Furthermore, the detailed
zenith angle and energy dependences observed for 
the atmospheric $\nu_{\mu}$ and $\nu_e$
fluxes are non-trivially consistent with a neutrino 
oscillation explanation based on either
$\nu_{\mu} \to \nu_{\tau}$ or $\nu_{\mu} \to \nu_s$\cite{fvy}
(where $\nu_s$ is a hypothetical
sterile neutrino). In this paper we will show that 
these results lead to a theoretical
explanation of the famous phenomenon of electric charge 
quantisation, provided that (i) the
mode responsible is $\nu_{\mu} \to \nu_{\tau}$, and (ii) 
that the neutrino mass involved is
of Majorana type.

The argument is extremely simple. We begin with the known 
result \cite{foot} that electric charge
quantisation (ECQ) does not follow from the Minimal Standard 
Model (MSM) Lagrangian (where
``minimal'' means zero neutrino mass). 
Therefore, new 
physics is required to explain it \cite{rev}. 
It is important to recall, however, that the possible 
charge dequantisation allowed by the
MSM is strongly constrained by the gauge invariance of the 
MSM Lagrangian and anomaly cancellation. 
Enforcing classical gauge invariance only, the allowed form 
for electric charge in the MSM is
\begin{equation}
Q_{\text{actual}} = Q_{\text{standard}} + \alpha L_e + \beta L_{\mu} 
+ \gamma L_{\tau} + \delta B,
\label{eq1}
\end{equation}
where $Q_{\text{actual}}$ and $Q_{\text{standard}}$ are 
the actual and standard electric
charges, $L_{e,\mu,\tau}$ are the three types of lepton number, 
and $B$ is baryon number. The quantities $L_{e,\mu,\tau}$ and $B$ enter this formula
precisely because they are conserved Abelian quantum numbers in the MSM.
The continuous parameters $\alpha$, $\beta$, $\gamma$ and 
$\delta$ quantify electric charge
dequantisation. The measured upper bounds on their magnitudes 
are, of course, tiny.
Enforcing gauge anomaly cancellation \cite{abj} in addition to 
classical gauge invariance leads to the
additional constraints
\begin{equation}
\gamma = (- \alpha^3 - \beta^3)^{1/3},\quad \delta 
= - \frac{1}{3} [\alpha + \beta +
(-\alpha^3 - \beta^3)^{1/3}],
\label{eq2}
\end{equation}
where there are now only two continuous charge dequantisation parameters 
$\alpha$ and $\beta$. If one chooses to
also enforce the cancellation of mixed gauge-gravitational 
anomalies\cite{ds}, then the
allowed electric charges are further constrained to be given by\cite{foot,rev}
\begin{eqnarray}
Q_{\text{actual}} & = & Q_{\text{standard}} 
+ \epsilon (L_e - L_{\mu}),\ \ \text{or}
\label{eq3a} \\
& = &  Q_{\text{standard}} + \epsilon (L_e - L_{\tau}),\ \ \text{or} 
\label{eq3b} \\
& = & Q_{\text{standard}} + \epsilon (L_{\mu} - L_{\tau}),
\label{eq3c}
\end{eqnarray}
where $\epsilon$ now quantifies charge dequantisation, in 
addition to the threefold
discrete choice one has between these forms. Note 
that baryon number $B$ plays no role if
mixed gauge-gravitational anomalies are forced to cancel. 
The fact that the totality of
classical and quantal constraints one may apply to 
the MSM do not uniquely specify
electric charge constitutes the modern statement of 
the electric charge quantisation problem.

It is interesting that the charge quantisation problem 
depends crucially on the
conservation of the family lepton numbers. Since neutrino 
oscillation experiments
search for family lepton number violation, they 
also indirectly probe the charge
quantisation problem. 

There are of course many experimental indications in favour 
of neutrino oscillations and
hence of family lepton number violation: the 
solar neutrino deficit \cite{solar} and the LSND anomaly \cite{lsnd}, in
addition to the aforementioned atmospheric neutrino deficit 
observed by SuperKamiokande and
other experiments \cite{others}. At present, the strongest experimental 
evidence for neutrino
oscillations comes from the SuperKamiokande atmospheric neutrino 
results. It is therefore
interesting to ask what one may conclude about 
the charge quantisation problem on the basis
of their results. Are the SuperKamiokande atmospheric neutrino 
results sufficient to
explain electric charge quantisation? If not, 
what further experimental information is needed?

Conservatively speaking, SuperKamiokande has established the 
violation of $L_{\mu}$. By
itself, this is not enough to explain ECQ, because charge 
may, for example, be dequantised
as per Eqn.(\ref{eq3b}) which does not involve $L_{\mu}$. As 
a concrete illustration, one
may explain the atmospheric neutrino results by $\nu_{\mu} \to 
\nu_s$ oscillations only,
leaving $\nu_e$ and $\nu_{\tau}$ as massless states unmixed 
with any others. Then, $L_e -
L_{\tau}$ is an anomaly-free conserved quantity that can 
dequantise electric charge. If one
does not wish to impose gauge and/or mixed anomaly cancellation 
then there are of course more possibilities.

So, let us now suppose that the atmospheric neutrino results 
are explained by $\nu_{\mu}
\to \nu_{\tau}$ oscillations. Since we know that $\nu_{\tau}$ almost certainly exists, 
this is perhaps a
less speculative (though not necessarily more attractive) 
possibility. Conservation of
$L_{\mu} + L_{\tau}$, $L_e$ and $B$ are consistent with this 
oscillation mode (but not
required by it). Oscillations of $\nu_{\mu}$ to $\nu_{\tau}$ 
require that at least one of
the mass eigenstates which superpose these flavour eigenstates to 
have a nonzero mass. (Of
course, in general one would expect all participating neutrino masses to 
be nonzero.)

The conclusions depend on whether this nonzero mass is of 
Majorana or Dirac type \cite{flv,bm}. If the 
mass is Dirac type, then the additional right-handed 
neutrino state required alters the
anomaly cancellation calculations which lead to Eqns.(\ref{eq2}) and
(\ref{eq3a}-\ref{eq3c}). 

Let us suppose first that it is of the more economical 
and theoretically more
appealing Majorana type. In this case, $L_{\mu}$ and $L_{\tau}$
must be separately broken. (This conclusion is obviously unaltered 
if there are more nonzero
neutrino eigenmasses.) One then immediately draws 
the interesting conclusion, either
from Eqns.(\ref{eq3a}-\ref{eq3c}) or from the more general 
Eqn.(\ref{eq2}), that electric charge
must be quantised in the standard way. If both $L_{\mu}$ and 
$L_{\tau}$ are broken, then
from the more general Eqn.(\ref{eq2}) we must set
\begin{equation}
\beta = 0, \quad \gamma = (-\alpha^3 - \beta^3)^{1/3} 
= 0\ \  \Rightarrow \ \ \alpha = 0,
\end{equation}
and hence $Q_{\text{actual}} = Q_{\text{standard}}$. {\it 
The SuperKamiokande atmospheric
neutrino data explain electric charge quantisation provided 
that $\nu_{\mu} \to \nu_{\tau}$
oscillations induced by a Majorana mass are their explanation.} 
This is our main result.
The only ways to evade it are either to give up gauge anomaly cancellation, or to suppose
that
there are as yet unobserved heavy fermions which contribute to anomaly cancellation 
\cite{classical}.
Note that both the choices -- of $\nu_{\mu} \to \nu_{\tau}$ 
over $\nu_{\mu} \to \nu_s$ and
of Majorana over Dirac mass -- are ones of minimality with 
respect to (low energy) degrees of freedom.

For completeness, we now discuss the Dirac mass alternative even 
though it is much less
appealing from a see-saw mechanism point of view. We 
first discuss the even less likely
supposition that there is only one nonzero Dirac 
mass. We will also make the conservative assumption that the $\nu_e$ does not mix the other
neutrinos.
Since $\nu_{\mu}$ and $\nu_{\tau}$ mix,
the single Dirac mass term must be of the form
\begin{equation}
m (\cos\theta \overline{\nu}_{\mu L} 
+ \sin\theta \overline{\nu}_{\tau L}) N_R + \text{H.c.}
\end{equation}
where $m$ is the mass, $\theta$ is the mixing 
angle\footnote{It is measured to be close to
$\pi/4$, though we only need to know that it is nonzero for 
the sake of our argument.} and
$N_R$ is a right-handed neutrino field. This mass 
term does not preserve $L_\mu$ or $L_\tau$
separately but does preserve $L_{\mu} + L_{\tau}$\footnote{
Provided of course that the $L_{\mu} + L_{\tau}$ charge of
$N_R$ is the same as the  $L_{\mu}+L_{\tau}$ charge
of the $\nu_\mu$ and $\nu_\tau$.} (as
well as, trivially, $L_e$ and $B$). Thus, 
the classical gauge invariance of the Lagrangian 
implies that
\begin{equation}
Q_{\text{actual}} = Q_{\text{standard}} + \alpha L_e + \beta (L_\mu + L_\tau)
+ \delta B.
\end{equation}
Futhermore, gauge anomaly cancellation only holds provided that
\begin{equation}
\beta = -\alpha = -3\delta.
\end{equation}
Hence in this case electric charge is not quantised, since
there is one theoretically unconstrained continuous parameter, 
$\epsilon = \alpha$, such that
\begin{equation}
Q_{\text{actual}} = Q_{\text{standard}} + \epsilon\left(
L_e - L_\mu - L_\tau + {B \over 3}\right).
\label{1nuR}
\end{equation}
Also note that in this case the mixed gauge gravitational anomaly
provides no independent constraint.

Suppose now that both the Dirac eigenmasses involving $\nu_{\mu}$ and 
$\nu_{\tau}$ are nonzero.
In this scenario there are two right-handed neutrino fields $N_{1R}$ 
and $N_{2R}$. The
quantities $L_{\mu} + L_{\tau}$, $L_e$ and $B$ 
are again classically conserved.
Following similar 
reasoning to the one right-handed neutrino scenario
discussed above, we find that electric charge
is also dequantised in this case, again with one theoretically unconstrained
continuous parameter, $\epsilon = \beta$, such that
\begin{equation}
Q_{\text{actual}} = Q_{\text{standard}} + \epsilon\left(
L_\mu + L_\tau -{2B \over 3}\right).
\label{2nuR}
\end{equation}
Again, the mixed gauge gravitational anomaly
provides no independent constraint in this case.

Equations (\ref{1nuR}) and (\ref{2nuR}) were derived on the assumption that $\nu_e$ does
not
mix with the other neutrinos. If we now alternatively suppose that it does mix (recall that
the LSND anomaly might be due to $\nu_e - \nu_{\mu}$ mixing), then $L_e$ and $L_{\mu} +
L_{\tau}$ are no longer separate invariances, with only the linear combination $L = L_e +
L_{\mu} + L_{\tau}$ (total lepton number) being conserved. The coefficient of $L_e$ in the
formula for electric charge must now be put equal to the coefficient of $L_{\mu} +
L_{\tau}$, meaning that $\epsilon = 0$ in Eqns.(\ref{1nuR}) and (\ref{2nuR}) and electric
charge quantisation now follows in both of these scenarios.

Finally, suppose that the most general and 
perhaps most likely situation holds (given the
Dirac mass assumption): there are three right-handed neutrino 
fields, and the neutrinos mix
in an arbitrary manner. In this case, $B-L$ is anomaly-free and 
so charge may be
dequantised as per $Q_{\text{actual}} = 
Q_{\text{standard}} + \epsilon (B-L)$\cite{rev,flv}.

To conclude, we have shown that the SuperKamiokande
atmospheric neutrino results explain the charge
quantisation mystery of the standard model provided that the
oscillation mode is $\nu_\mu \to \nu_\tau$ and the
the neutrino mass is of the Majorana type. It is important in general to realise
that neutrino oscillation and
neutrinoless double $\beta$-decay experiments provide important information regarding the
seemingly unrelated issue of electric charge quantisation.
 
\acknowledgments{This work was supported by the Australian Research
Council. R.F. is an Australian Research Fellow. We thank Yvonne Wong for some helpful
advice.}

\end{document}